\documentclass[%
 reprint,showkeys,
onecolumn,superscriptaddress,
bibnotes,
 amsmath,amssymb,
 aps,
pra,
]{revtex4-2}

\usepackage{graphicx}
\usepackage{dcolumn}
\usepackage{physics}
\usepackage{bm}
\usepackage{hyperref}

\begin{document}

\preprint{APS/123-QED}

\title{Rydberg atom-based microwave electrometry using polarization spectroscopy}

\author{Naomy Duarte Gomes}
 \affiliation{Instituto de F\'{i}sica de S\~{a}o Carlos, Universidade de S\~{a}o Paulo, Caixa Postal 369, 13560-970, S\~{a}o Carlos, SP, Brasil.}
\author{Vinicius Marrara Pepino}%
 \affiliation{Departamento de Engenharia Elétrica e Computação, Escola de Engenharia de São Carlos, Universidade de São Paulo, São Carlos, Brasil
}%
\author{Ben-Hur Viana Borges}
 \affiliation{Departamento de Engenharia Elétrica e Computação, Escola de Engenharia de São Carlos, Universidade de São Paulo, São Carlos, Brasil
}%
\author{Daniel Varela Magalhães}
 \affiliation{Instituto de F\'{i}sica de S\~{a}o Carlos, Universidade de S\~{a}o Paulo, Caixa Postal 369, 13560-970, S\~{a}o Carlos, SP, Brasil.}
\author{Reginaldo de Jesus Napolitano}
 \affiliation{Instituto de F\'{i}sica de S\~{a}o Carlos, Universidade de S\~{a}o Paulo, Caixa Postal 369, 13560-970, S\~{a}o Carlos, SP, Brasil.}
 \author{Manuel Alejandro Lefr{\'a}n Torres}
 \affiliation{Instituto de F\'{i}sica de S\~{a}o Carlos, Universidade de S\~{a}o Paulo, Caixa Postal 369, 13560-970, S\~{a}o Carlos, SP, Brasil.}
 \author{Jorge Douglas Massayuki Kondo}
 \affiliation{Departamento de Física, Universidade Federal de Santa Catarina, Florianópolis 88040-900, SC, Brasil}
  \author{Luis Gustavo Marcassa}
   \email{marcassa@ifsc.usp.br}
 \affiliation{Instituto de F\'{i}sica de S\~{a}o Carlos, Universidade de S\~{a}o Paulo, Caixa Postal 369, 13560-970, S\~{a}o Carlos, SP, Brasil.}

\date{\today}

\begin{abstract}
In this study, we investigated Rydberg atom-based microwave electrometry using polarization spectroscopy in a room-temperature vapor cell. By measuring Autler-Townes splitting in the electromagnetically induced transparency (EIT) spectrum, we determined that the minimum measurable microwave electric field is approximately five times lower than conventional EIT techniques. The results are well reproduced by a full optical Bloch equation model, which takes into account all the hyperfine levels involved. Subsequently, the EIT setup was used to characterize a custom microwave cylindrical lens, which increases the field at the focus by a factor of three, decreasing the minimum measurable microwave electric field by the same amount. Our results indicate that the combination of polarization spectroscopy and a microwave lens may enhance microwave electrometry.
\end{abstract}

\maketitle

\section{\label{sec:level1}Introduction}

Polarization spectroscopy (PS) is a powerful technique for studying atomic and molecular spectra. Medium birefringence is induced by a circularly polarized pump beam and is interrogated with a counterpropagating weak probe beam \cite{Wieman76}. This technique has demonstrated its effectiveness in spectroscopy in various fields, including laser frequency stabilization \cite{Pearman2002}, plasma characterization \cite{Danzmann1986}, and even combustion diagnostics \cite{Combustion}. 

\begin{figure}
      \centering
    \includegraphics[width=0.6\columnwidth]{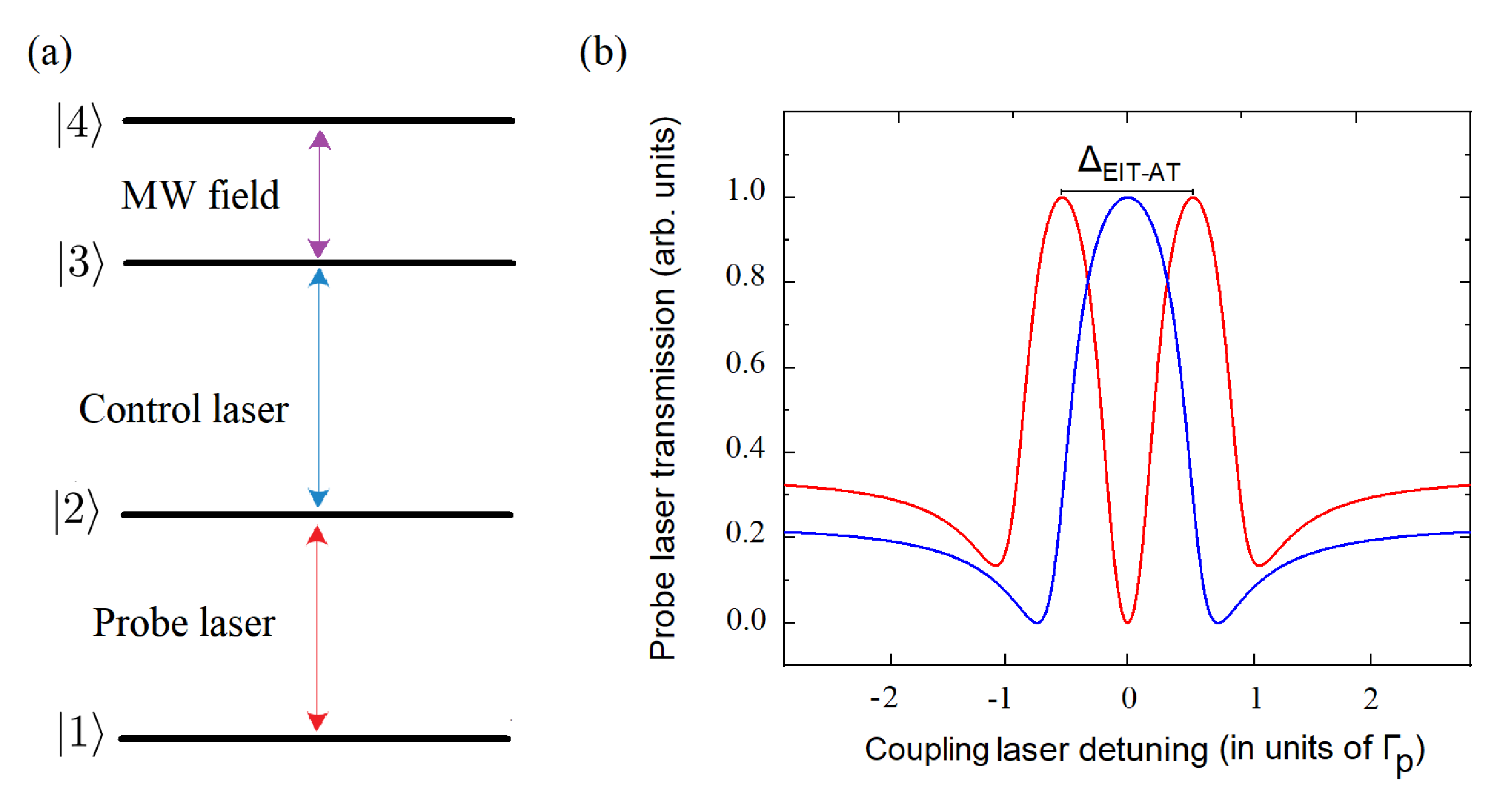}
    \caption{(Color online) (a) Scheme of the four-level atom. Levels $\ket{1}$ and $\ket{2}$ are coupled by a probe laser, levels $\ket{2}$ and $\ket{3}$ are coupled by a coupling laser, and levels $\ket{3}$ and $\ket{4}$ are Rydberg levels connected by MW radiation. (b) The EIT transmission spectrum with no MW field (blue) and with an applied MW field (red), evidencing the $\Delta_{EIT-AT}$, in units of the decaying rate $\Gamma_{p}$ from state $\ket{2}$.}
    \label{fig01}
\end{figure}

Electromagnetically induced transparency (EIT) is a nonlinear optical effect found in three-level atomic systems \cite{Marangos1998,Marangos2005}. In this phenomenon, a medium normally opaque to a weak probe electromagnetic field can turn transparent when influenced by a strong coupling electromagnetic field. This effect is marked by changes in the absorption profile and a notable decrease in light velocity \cite{Hau1999}, thereby extending the duration of the interaction between light and matter. EIT has been used to investigate Rydberg atoms \cite{Clack2001,mohapatra2007}, leading to the development of Rydberg atom-based microwave (MW) electrometry \cite{sedlacek2012microwave}. This technique is based on a four-level atom excited in a ladder configuration, as shown in Fig. \ref{fig01} (a). Levels $\ket{1}$ and $\ket{2}$ are coupled by a probe laser, levels $\ket{2}$ and $\ket{3}$ are coupled by a coupling laser, and levels $\ket{3}$ and $\ket{4}$ are Rydberg levels connected by MW radiation. In the regime of weak/strong probe/coupling lasers, the EIT transparency spectrum presents an Autler-Townes (AT) splitting \cite{cohen1996autler, anisimov2011objectively, abi2010electromagnetically} (Fig. \ref{fig01} (b), $\Delta_{EIT-AT}$), whose depth and shape are sensitive to the strength and frequency of the MW field. $\Delta_{EIT-AT}$ is a direct and precise measurement of the electric field strength of MW and is traceable to the International System of Units (SI) \cite{sedlacek2012microwave}.

\indent So far polarization spectroscopy has been applied to Rydberg EIT only for laser frequency stabilization \cite{abel2009,carr2012polarization, meyer2017nonlinear} and for line narrowing using a Laguerre–Gauss coupling laser beam \cite{duarte2022polarization}. In this work, we have investigated Rydberg atom-based microwave electrometry using polarization spectroscopy in an EIT spectrum using a room-temperature vapor cell, which we call PSEIT. Measurements of Autler-Townes splitting ($\Delta_{PSEIT-AT}$), within the PSEIT signal, demonstrate that the minimum detectable microwave electric field is roughly five times smaller compared to traditional EIT methods \cite{sedlacek2012microwave}. A comprehensive optical Bloch equation model, considering all relevant hyperfine levels, successfully replicates these findings. Subsequently, the EIT setup was used to characterize a specialized MW cylindrical lens, which triples the field strength at its focal point, thereby tripling the reduction in the minimum detectable microwave electric field. Our findings suggest that the use of both polarization spectroscopy and a microwave lens may improve microwave electrometry by $\simeq 15$.

The structure of this work is organized as follows: i) Section  \ref{exp-setup} presents the experimental setup and procedure; ii) Section \ref{th} provides the theoretical model; iii) Section \ref{exp-res} provides the experimental results and discussion, which includes the PSEIT-AT results (Subsection  \ref{pseit-at}), the design and simulation of the MW lens (Subsection \ref{sec:lens-design}), and the MW lens results (Subsection \ref{lens}); iv) Finally, Section \ref{conclu} concludes the study with the presentation of the main findings and conclusions.

\section{Experimental setup and procedure}\label{exp-setup}

\begin{figure}
    \centering
    \includegraphics[width=0.6\columnwidth]{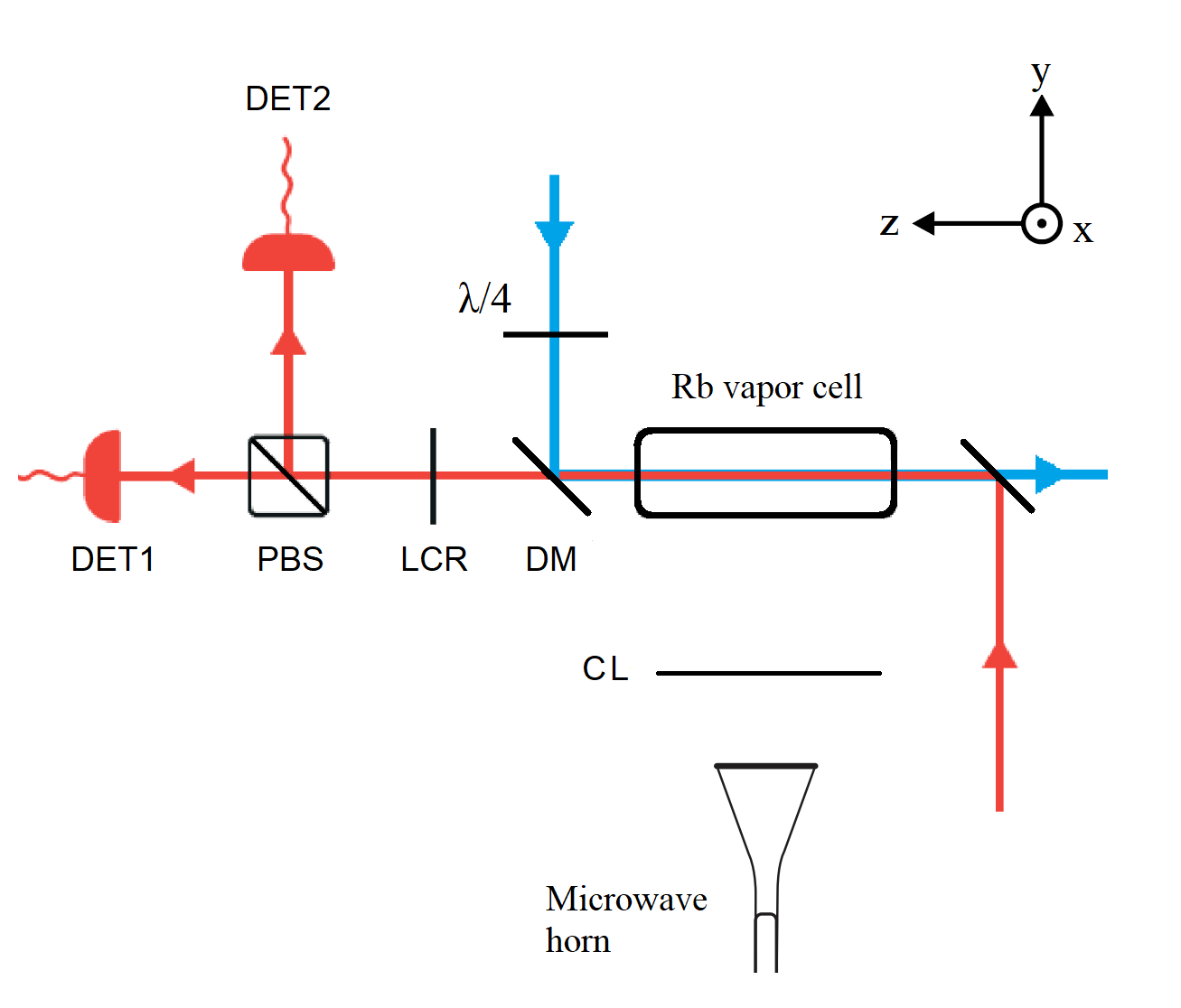}      
    \caption{(Color online) Experimental setup (not to scale):  waveplate ($\lambda/4$); polarizing beam splitter (PBS); Dichroic mirror (DM); liquid crystal retarder (LCR);  MW cylindrical lens (CL); vapor cell and photodetectors (DET1 and DET2).  A probe (red) and coupling (blue) laser beams counterpropagate inside a rubidium vapor cell under the influence of an MW field generated at the MW horn antenna. The probe laser beam is linearly polarized in the x-direction, while the MW electric field is polarized in the z-direction. For the EIT measurement, the coupling laser beam is linear polarized in the x-direction. For the PSEIT measurement, the $\lambda/4$ is used to make it circular polarized. The probe beam is further detected at DET1 and DET2.}
    \label{fig02}
\end{figure}

Figure \ref{fig01}(a) shows the four-level atomic ladder configuration associated with the EIT effect ($\ket{1}=5S_{1/2}$, $\ket{2}=5P_{3/2}$, $\ket{3}=68S_{1/2}$, and $\ket{4}=68P_{3/2}$). Details of the experimental arrangement are reported in \cite{duarte2022polarization} and depicted in Fig. \ref{fig02}. In summary, the probe (red line) and coupling (blue line) laser beams travel in opposite directions within a 7.5 cm Rb vapor cell maintained at ambient temperature. A dichroic mirror separates and analyzes the probe beam. Both beams are passed through single-mode optical fibers to achieve a Gaussian profile and are focused at the cell's center, which lacks magnetic shielding. The probe laser beam operates at a total power of 4.5 $\mu$W, with a $1/e^{2}$ waist of 170 $\pm$ 5 $\mu$m (Rayleigh length $z_{R} \approx 116$ mm), corresponding to a calculated $\Omega_{p}/2\pi = 2.32$ MHz. This laser is stabilized to an optical cavity for the $5S_{1/2}(F=3) \rightarrow 5P_{3/2} (F=4)$ transition $^{85}Rb$ with a sub $0.1$ MHz linewidth at 780 nm \cite{RodriguezFernandez2023}. The coupling laser beam, with a power of 36 mW and a $1/e^2$ waist of $190 \pm 5$ $\mu$m (Rayleigh length $z_{R} \approx 236$ mm), has a calculated Rabi frequency of $\Omega_{c}/2\pi = 3.56$ MHz. It scans over the $5P_{3/2}(F=4) \rightarrow 68S_{1/2} (F=3)$ transition. A microwave signal generator connected to a horn antenna (Model PE9856B/SF-15 Pasternack) excites the $68S_{1/2} \rightarrow 68P_{3/2}$ transition at 11.6662 $\pm$ 0.0001 GHz. The MW horn antenna is placed 82 cm from the vapor cell. When the MW cylindrical lens (CL) is used, it is placed between the horn antenna and the vapor cell, at 27 cm from the latter. After the dichroic mirror, the probe beam passes through a half-wave liquid crystal retarder (LCR) (Thorlabs model LCC1111-B) and a polarizing beam splitter (PBS), which decomposes it into two orthogonal linearly polarized beams. Each beam is detected by a photodiode (DET1 and DET2). 

To detect the EIT signal, the probe and the coupling laser beams are linearly polarized in the x-direction, while the MW electric field is polarized in the z-direction and a single detector detects the probe laser. To detect the PSEIT signal, both the probe and microwave fields are linearly polarized as before, while the coupling laser beam is circularly polarized. In this case, the photodiode signals are subtracted, resulting in a dispersive signal that is due to the anisotropy generated in the atomic medium by the coupling beam \cite{carr2012polarization, meyer2017nonlinear}.  This dispersive signal is the PSEIT signal. The LCR is ideal for this task because it allows a precise balance of the final signal since it does not have moving parts. To increase the signal-to-noise ratio, the coupling beam is modulated at 4 kHz, and the detected signal is processed by a lock-in amplifier.

\section{Theoretical Model}\label{th}

Polarization spectroscopy has been theoretically investigated using optical Bloch equations for the D2 transitions in Rb and Cs \cite{Harris2006}. Here we extend such a model for an EIT configuration, where we explicitly calculate the ground and excited state probabilities and coherences. First, we calculate the density operator and then the dispersion profile.

\subsection{The density operator}

The atomic Hamiltonian, $H_{0},$ can be expressed in terms of projectors,
$\Pi_{g},$ $\Pi_{e},$ $\Pi_{g_{Q}},$ $\Pi_{e_{Q1}},$ and $\Pi_{e_{Q2}},$
which will be described shortly:
\begin{eqnarray}
H_{0} & = & \hbar\omega_{0}\Pi_{g}+\hbar\omega_{e}\Pi_{e}+\hbar\omega_{g_{Q}}\Pi_{g_{Q}} +\hbar\omega_{e_{Q1}}\Pi_{e_{Q1}}+\hbar\omega_{e_{Q2}}\Pi_{e_{Q2}},
\end{eqnarray}
where $\hbar\omega_{0},$ $\hbar\omega_{e},$ $\hbar\omega_{g_{Q}},$
$\hbar\omega_{e_{Q_{1}}},$ and $\hbar\omega_{e_{Q2}}$ are the respective
energies of the states in the $5^{2}S_{1/2}\left(F=3\right),$ $5^{2}P_{3/2}\left(F=4\right),$
$68^{2}S_{1/2}\left(F=3\right),$ $68^{2}P_{3/2}\left(F=3\right),$
and $68^{2}P_{3/2}\left(F=4\right)$ manifolds. For simplicity we
choose $\omega_{0}=0.$ As we can see in the expression of $H_{0},$
each of the $5^{2}S_{1/2}\left(F=3\right),$ $5^{2}P_{3/2}\left(F=4\right),$
$68^{2}S_{1/2}\left(F=3\right),$ $68^{2}P_{3/2}\left(F=3\right),$
and $68^{2}P_{3/2}\left(F=4\right)$ manifolds is degenerate, in accordance
with our assumption that magnetic fields are negligible.

Thus, the model we use consists of $40$ states in total, one of which,
state $\left|D\right\rangle ,$ is an auxiliary state, artificially
introduced to account for the transit-time broadening. For simplicity,
we also take the artificial state $\left|D\right\rangle $ to have
zero energy. The $39$ non-artificial atomic states are the hyperfine
magnetic states defined relative to the $z$ axis shown in Fig. \ref{fig02}. The $40\mathrm{D}$ Hilbert space we use can be analyzed in subspaces.
Let $P$ be the projector operator that projects onto the $16\mathrm{D}$
subspace of the $5^{2}S_{1/2}\left(F=3\right)$ and $5^{2}P_{3/2}\left(F=4\right)$
manifolds. Accordingly, let $Q=I-P$ be the projector onto the
$24\mathrm{D}$ subspace of the $68^{2}S_{1/2}\left(F=3\right)$ and
$68^{2}P_{3/2}\left(F=3,4\right)$ manifolds, together with the artificial
state $\left|D\right\rangle .$

The lower manifold, represented by $P,$ can be split into two subspaces:
the $5^{2}S_{1/2}\left(F=3\right),$ with a $7\mathrm{D}$ projector
$\Pi_{g},$ and the $5^{2}P_{3/2}\left(F=4\right),$ with a $9\mathrm{D}$
projector $\Pi_{e},$ so that $P=\Pi_{g}+\Pi_{e}.$ The Liouvillian
term in the master equation, used to include the spontaneous decay
from the upper to the lower hyperfine magnetic states in the $P$
manifold, is given by
\begin{eqnarray}
L_{rad}\left[\tilde{\rho}\left(t\right)\right] & = & \Pi_{g}\boldsymbol{\Gamma}\tilde{\rho}\left(t\right)\boldsymbol{\cdot}\mathbf{d}\Pi_{g}-\Pi_{g}\mathbf{d}\tilde{\rho}\left(t\right)\boldsymbol{\cdot}\boldsymbol{\Gamma}\Pi_{g}\nonumber \\
 &  & -\frac{1}{2}\left\{ \Pi_{e}\mathbf{d}\boldsymbol{\cdot}\boldsymbol{\Gamma}\Pi_{e}-\Pi_{e}\boldsymbol{\Gamma}\boldsymbol{\cdot}\mathbf{d}\Pi_{e},\tilde{\rho}\left(t\right)\right\} ,
\end{eqnarray}
where $\tilde{\rho}\left(t\right)$ is the reduced atomic density
operator relative to the ``rotating'' frame of reference, $\mathbf{d}$
is the atomic electric dipole moment operator, and $\boldsymbol{\Gamma}$
is the operator defined by
\begin{eqnarray}
\boldsymbol{\Gamma} & = & \frac{2\omega_{e}^{3}}{3\hbar c^{3}}\left[\mathbf{d},\Pi_{e}\right],
\end{eqnarray}
where $\omega_{e}$ is the degenerate energy of the $5^{2}P_{3/2}\left(F=4\right)$
magnetic states. Here we neglect magnetic fields and choose the degenerate
energy of the $5^{2}S_{1/2}\left(F=3\right)$ states as zero.

The $Q$ manifold consists of the $68^{2}S_{1/2}\left(F=3\right)$
subspace, with a $7\mathrm{D}$ projector $\Pi_{g_{Q}},$ the $68^{2}P_{3/2}\left(F=3\right)$
subspace, with another $7\mathrm{D}$ projector $\Pi_{e_{Q1}},$ the
$68^{2}P_{3/2}\left(F=4\right)$ subspace, with a $9\mathrm{D}$ projector
$\Pi_{e_{Q2}},$ and the $\left|D\right\rangle $ subspace, with a
$1\mathrm{D}$ projector $\Pi_{D}=\left|D\right\rangle \left\langle D\right|,$
which is the artificial subspace we use to treat the transit-time
decay. We do not consider the spontaneous emission from the $Q$ states,
since their radioactive lifetime is longer than the short-time transit
to cross the light beam. In the master equation, the transit-time decay is included using a Lindbladian term given by
\begin{eqnarray}
L_{t}\left[\tilde{\rho}\left(t\right)\right] & = & \gamma\left\langle D\right|\tilde{\rho}\left(t\right)\left|D\right\rangle \Pi_{g}-\frac{7}{2}\gamma\left\{ \Pi_{D},\tilde{\rho}\left(t\right)\right\}+\gamma\Pi_{D}\mathrm{Tr}\left[\Pi_{g_{Q}}\tilde{\rho}\left(t\right)\right] \nonumber \\
 &  & -\frac{\gamma}{2}\left\{ \Pi_{g_{Q}},\tilde{\rho}\left(t\right)\right\}+\gamma\Pi_{D}\mathrm{Tr\left[\Pi_{e_{Q}}\tilde{\rho}\left(t\right)\right]}-\frac{\gamma}{2}\left\{ \Pi_{e_{Q}},\tilde{\rho}\left(t\right)\right\} ,
\end{eqnarray}
where $\gamma$ is the effective transit-time decay rate for all $n=68$ states and $\Pi_{e_{Q}}=\Pi_{e_{Q1}}+\Pi_{e_{Q2}}.$

We take $\omega_{p},$ $\omega_{c},$ and $\omega_{m}$ to stand for
the frequencies of the probe, coupling, and microwave laser fields,
respectively. The coefficients in the resulting master equation are
made time-independent after a unitary transformation $\exp\left(-iTt\right),$
where
\begin{eqnarray}
T & = & \omega_{p}\Pi_{e}+\left(\omega_{c}+\omega_{p}\right)\Pi_{g_{Q}} +\left(\omega_{m}+\omega_{c}+\omega_{p}\right)\Pi_{e_{Q}},
\end{eqnarray}
so that
\begin{eqnarray}
\tilde{\rho}\left(t\right) & = & \exp\left(iTt\right)\rho\left(t\right)\exp\left(-iTt\right),
\end{eqnarray}
where $\rho\left(t\right)$ stands for the reduced density operator
in the interaction picture, before the unitary transformation just
mentioned. We also define the detunings:
\begin{eqnarray}
\Delta_{pe} & = & \omega_{p}-\omega_{e},\\
\Delta_{cg_{Q}} & = & \omega_{c}-\omega_{g_{Q}}+\omega_{e},\\
\Delta_{me_{Q1}} & = & \omega_{m}-\omega_{e_{Q1}}+\omega_{g_{Q}},\\
\Delta_{me_{Q2}} & = & \omega_{m}-\omega_{e_{Q2}}+\omega_{g_{Q}}.
\end{eqnarray}

Relatively to the axes in Fig. \ref{fig02}, the probe electric field is linearly
polarized along the $x$ axis, the coupling electric field has $\sigma^{+}$ polarization, and the microwave electric field is polarized along
the $z$ axis. Accordingly, using the versors,
\begin{eqnarray}
\boldsymbol{\hat{\varepsilon}}_{\pm1} & = & \mp\left(\frac{\mathbf{\hat{x}}\pm i\mathbf{\hat{y}}}{\sqrt{2}}\right)\label{epsilonpm}
\end{eqnarray}
and
\begin{eqnarray}
\boldsymbol{\hat{\varepsilon}}_{0} & = & \mathbf{\hat{z}},
\end{eqnarray}
the Rabi operators for the probe, coupling, and microwave laser fields
are, respectively:
\begin{eqnarray}
\hbar\Omega_{p} & = & -\Pi_{g}\mathbf{d}\boldsymbol{\cdot}\boldsymbol{\hat{\varepsilon}}_{+1}^{\ast}\Pi_{e}\frac{E_{p}}{\sqrt{2}}+\Pi_{g}\mathbf{d}\boldsymbol{\cdot}\boldsymbol{\hat{\varepsilon}}_{-1}^{\ast}\Pi_{e}\frac{E_{p}}{\sqrt{2}},
\end{eqnarray}
\begin{eqnarray}
\hbar\Omega_{c} & = & \Pi_{e}\mathbf{d}\boldsymbol{\cdot}\boldsymbol{\hat{\varepsilon}}_{+1}^{\ast}\Pi_{g_{Q}}E_{c},
\end{eqnarray}
and
\begin{eqnarray}
\hbar\Omega_{m} & = & \Pi_{g_{Q}}\mathbf{d}\boldsymbol{\cdot}\boldsymbol{\hat{\varepsilon}}_{0}\Pi_{e_{Q}}E_{m},
\end{eqnarray}
where $E_{p},$ $E_{c},$ and $E_{m}$ are the probe, coupling, and
microwave electric field amplitudes, respectively. The Hamiltonian
describing the interaction between the atom and the laser fields is
thus given by
\begin{eqnarray}
H_{int}\left(0\right) & = & -\frac{\hbar}{2}\left(\Omega_{p}^{\dagger}+\Omega_{p}\right)-\frac{\hbar}{2}\left(\Omega_{c}^{\dagger}+\Omega_{c}\right) -\frac{\hbar}{2}\left(\Omega_{m}^{\dagger}+\Omega_{m}\right).
\end{eqnarray}

We can now write the complete master equation for our multi-state
model as
\begin{eqnarray}
\frac{d}{dt}\tilde{\rho}\left(t\right) & = & -\frac{i}{\hbar}\left[H_{0}-\hbar T+H_{int}\left(0\right),\tilde{\rho}\left(t\right)\right] +L_{rad}\left[\tilde{\rho}\left(t\right)\right]+L_{t}\left[\tilde{\rho}\left(t\right)\right].\label{mstreq}
\end{eqnarray}
In our model, we are only considering the atoms that are, at any given moment, inside the laser interaction region. Assuming a uniform atomic density in the vapor cell, it is logical that for every atom exiting this region, another atom enters, maintaining a constant average population of atoms in both excited and ground states, and therefore in an average steady state. Accordingly, we solve the master equation, Eq. (\ref{mstreq}), from $t=0$ until $d\tilde{\rho}\left(t\right)/dt\approx0,$
reaching the steady-state solution for the density matrix operator.
Specifically, we make sure that, for each run, the time evolution
is stopped only after each density matrix element becomes constant
up to six decimal figures. We found that, for this $40\times40$ matrix
problem, the time integration to reach a steady state, as just specified,
is faster than imposing $d\tilde{\rho}\left(t\right)/dt=0$ and solving
the resulting linear system of equations.

\subsection{The dispersion profile}

The polarization operator is given by
\begin{eqnarray}
\mathbf{P}\left(t\right) & = & N\mathbf{d}\left(t\right),\label{polarization}
\end{eqnarray}
where $N$ is the atomic density. The complex version of the probe electric field at the entrance of the vapor cell, at
$z=0,$ is given by
\begin{eqnarray}
\mathbf{E}_{p}\left(0,t\right) & = & \mathbf{\hat{x}}\frac{E_{0}}{2}\exp\left(-i\omega_{p}t\right).
\end{eqnarray}
Upon interaction with the atoms, the electric field within the vapor
cell, for $0<z<L,$ where $L$ is the cell longitudinal length along
the positive $z\text{-axis},$ propagates differently depending on
the polarization state we consider. Using Eq. (\ref{epsilonpm}) we
can rewrite the complex probe electric field propagating in the atomic sample
as
\begin{eqnarray}
\mathbf{E}_{p}\left(z,t\right) & = & \frac{E_{0}}{2\sqrt{2}}\left[-\boldsymbol{\hat{\varepsilon}}_{+1}\exp\left(ik_{+}z-i\omega_{p}t\right) +\boldsymbol{\hat{\varepsilon}}_{-1}\exp\left(ik_{-}z-i\omega_{p}t\right)\right],\label{ansatz}
\end{eqnarray}
where the wave numbers $k_{\pm}$ indicate the different dispersion
characteristics that are dependent on the polarization state.

The electric dipole moment operator, because it does not couple states
of equal parity, is written as
\begin{eqnarray}
\mathbf{d} & = & \sum_{k}\sum_{e_{k}}\left(\left|k\right\rangle \mathbf{d}_{ke_{k}}\left\langle e_{k}\right|+\left|e_{k}\right\rangle \mathbf{d}_{e_{k}k}\left\langle k\right|\right),
\end{eqnarray}
where $k$ labels all the states that are excitable by the lasers
present in the experiment and $e_{k}$ labels, for each $k,$ all
the excited states that are reachable from the $k\text{th}$ state
by whatever resonant photons are present. In the Heisenberg picture,
the expected value of the electric dipole moment operator, already
in the rotating reference frame, gives
\begin{eqnarray}
\left\langle \mathbf{d}\left(z,t\right)\right\rangle  & = & \sum_{g,e}\left[\tilde{\rho}_{eg}\left(z,t\right)\mathbf{d}_{ge}\exp\left(-i\omega_{p}t\right) +\tilde{\rho}_{ge}\left(z,t\right)\mathbf{d}_{eg}\exp\left(i\omega_{p}t\right)\right],
\end{eqnarray}
where now we explicit the dependence on position along the $z\text{-axis}.$
Here, because, in the rotating reference frame, we use the rotating-wave
approximation, only the $5S_{1/2}\left(F=3\right)$ ground, $\left|g\right\rangle ,$
and $5P_{3/2}\left(F=4\right)$ excited, $\left|e\right\rangle ,$
states contribute appreciably.

Assuming the electric dipole matrix elements, $\mathbf{d}_{ge}=\mathbf{d}_{eg},$
to be real vectors, we can decompose them in their spherical tensor
components:
\begin{eqnarray}
\mathbf{d}_{ge} & = & -\boldsymbol{\hat{\varepsilon}}_{+1}\left\langle g\right|d_{-1}\left|e\right\rangle -\boldsymbol{\hat{\varepsilon}}_{-1}\left\langle g\right|d_{+1}\left|e\right\rangle  +\mathbf{\hat{z}}\left\langle g\right|d_{0}\left|e\right\rangle .\label{deg}
\end{eqnarray}
Hence, we can write the expected value of the complex version of the
polarization operator, Eq. (\ref{polarization}), in the rotating-frame
Heisenberg picture, as
\begin{eqnarray}
\mathbf{P}\left(z,t\right) & = & N\sum_{g,e}\tilde{\rho}_{eg}\left(z,t\right)\mathbf{d}_{ge}\exp\left(-i\omega_{p}t\right),
\end{eqnarray}
with $\mathbf{d}_{ge}$ expressed as in Eq. (\ref{deg}).

To obtain the dispersion characteristics, we now use the complex version
of Maxwell equations. In the steady state regime, where $\partial\tilde{\rho}_{eg}\left(z,t\right)/\partial t=0,$
we write only $\tilde{\rho}_{eg}\left(z\right).$ Then, using Maxwell
equations and the ansatz expressed by
Eq. (\ref{ansatz}), we obtain
\begin{eqnarray}
k_{+}^{2} & = & \frac{\omega_{p}^{2}}{c^{2}}\left[1+f\sum_{g,e}\frac{\tilde{\rho}_{eg}\left(z\right)}{\exp\left(ik_{+}z\right)}\left\langle g\right|d_{-1}\left|e\right\rangle \right]
\end{eqnarray}
and
\begin{eqnarray}
k_{-}^{2} & = & \frac{\omega_{p}^{2}}{c^{2}}\left[1-f\sum_{g,e}\frac{\tilde{\rho}_{eg}\left(z\right)}{\exp\left(ik_{-}z\right)}\left\langle g\right|d_{+1}\left|e\right\rangle \right],
\end{eqnarray}
where
\begin{eqnarray}
    f=\frac{8\pi\sqrt{2}N}{E_{0}}.
\end{eqnarray}

Rewriting Eq. (\ref{ansatz})
in terms of its components along the $x$ and $y$ axes, the probe electric
field, at the end of its propagation within the vapor cell, at $z=L,$ is given by
\begin{eqnarray}
\mathbf{E}_{p}\left(L,t\right) & = & \frac{E_{0}}{4}\mathbf{\hat{x}}\exp\left(ik_{+}L-i\omega_{p}t\right) +\frac{E_{0}}{4}\mathbf{\hat{y}}i\exp\left(ik_{+}L-i\omega_{p}t\right)\nonumber \\
 &  & +\frac{E_{0}}{4}\mathbf{\hat{x}}\exp\left(ik_{-}L-i\omega_{p}t\right) -\frac{E_{0}}{4}\mathbf{\hat{y}}i\exp\left(ik_{-}L-i\omega_{p}t\right).\label{Ec}
\end{eqnarray}
After leaving the vapor cell, the probe laser field continues its
propagation along the $z$ axis until it reaches the dichroic mirror,
DM, traverses it, and reaches, at a point that we will call $z_{1},$ the
position where it enters the liquid crystal retarder, LCR (see Fig. \ref{fig02} for setup details). After leaving the LCR, the $y$ component, let us say, gets multiplied by $-1,$ so that, for $z>z_{1},$ the probe laser
field can now be approximately described by
\begin{eqnarray}
\mathbf{E}_{p2}\left(z,t\right) & = & \mathbf{E}_{p1}\exp\left[ik_{0}\left(z-z_{1}+L\right)-i\omega_{p}t\right],
\end{eqnarray}
where $k_{0}$ is the vacuum wave number and $\mathbf{E}_{p1}$
is obtained from Eq. (\ref{Ec}) by changing $\mathbf{\hat{y}}$ into
$-\mathbf{\hat{y}},$ namely,
\begin{eqnarray}
\mathbf{E}_{p1} & = & \frac{E_{0}}{4}\exp\left(-k_{I+}L\right)\left(\mathbf{\hat{x}}-i\mathbf{\hat{y}}\right)\exp\left(ik_{R+}L\right) \nonumber \\
 &  & +\frac{E_{0}}{4}\exp\left(-k_{I-}L\right)\left(\mathbf{\hat{x}}+i\mathbf{\hat{y}}\right)\exp\left(ik_{R-}L\right),
\end{eqnarray}
where now we write
\begin{eqnarray}
k_{\pm} & = & k_{R\pm}+ik_{I\pm},
\end{eqnarray}
with $k_{R \pm}, k_{I \pm} \in \mathbb{R}.$

At the position of the polarizing beam splitter, PBS (see Fig. 2),
the $x$ and $y$ electric field components are separated and we can
calculate their time-averaged intensities. The dispersion profile ($S\left(\Delta\right)$)
is obtained by subtracting these $x\text{-}$ and $y\text{-component}$
signals:
\begin{eqnarray}
S\left(\Delta\right) & = & A\exp\left[\left(-k_{I+}-k_{I-}\right)L\right]\mathrm{sen}\theta\cos\theta,
\end{eqnarray}
where $A$ is an arbitrary amplitude and we have defined
\begin{eqnarray}
\theta & = & \frac{\pi}{\lambda_{p}}\left(n_{+}-n_{-}\right)L,
\end{eqnarray}
where
\begin{eqnarray}
n_{\pm} & = & \frac{c}{\omega_{p}}k_{R\pm}
\end{eqnarray}
are the refractive indexes and $\lambda_{p}=2\pi c/\omega_{p}.$ Finally, the dispersion profile is velocity averaged using a Maxwell-Boltzmann distribution at room temperature.

\section{Experimental Results and Discussion}\label{exp-res}

\subsection{EIT and PSEIT results}\label{pseit-at}

\begin{figure}
    \centering
    \includegraphics[width=0.6\columnwidth]{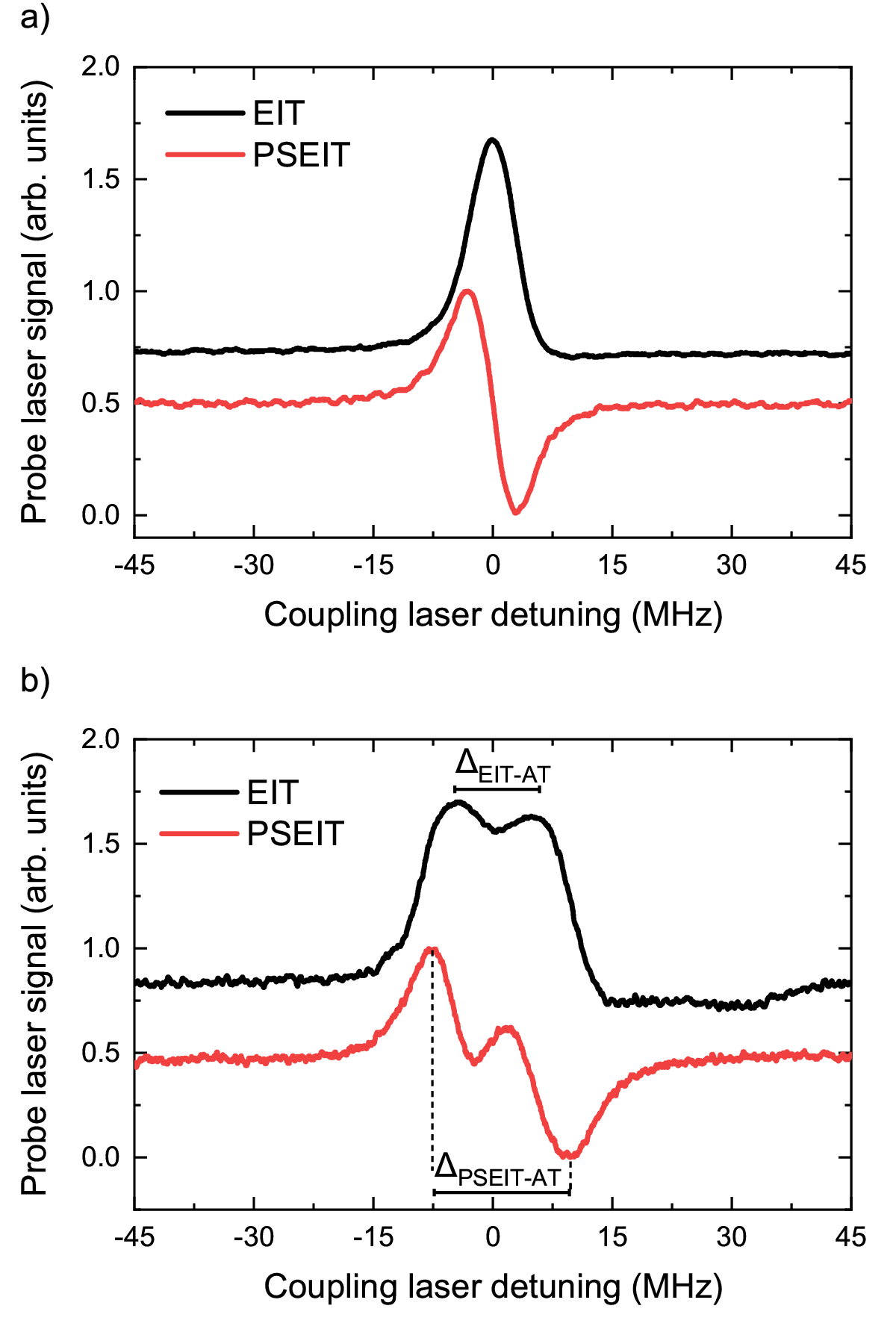}
    \caption{(Color online) (a) Typical EIT (black) and PSEIT (red) as a function of coupling laser detuning when no MW is applied. (b) EIT (black) and PSEIT (red) spectra as a function of coupling laser detuning when an MW field is applied (-4 dBm power). $\Delta_{EIT-AT}$ ($\Delta_{PSEIT-AT}$) is the MW field splitting for the EIT-AT (PSEIT-AT) signal.}
    \label{fig03}
\end{figure}

Figure \ref{fig03} shows typical EIT and PSEIT spectra as functions of coupling laser detuning. In Fig. \ref{fig03} (a), the EIT (black) and PSEIT (red) signals are obtained when no MW is present, while in Fig. \ref{fig03} (b) the EIT transmission signal (in black) shows two peaks, known as Autler-Townes doublets, separated by $\Delta_{EIT-AT}$, which is proportional to the strength of the applied MW field \cite{sedlacek2012microwave}. While one might assume that calculating the numerical derivative of the EIT signal would yield results similar to those of the PSEIT signal, our attempts at this approach proved otherwise. The numerical derivative of the EIT signal was significantly noisier when compared to the PSEIT signal, leading us to discard this method.

For a fair comparison of both techniques, we processed both signals identically. We obtained $\Delta_{EIT-AT}$  through signal  processing using the \textit{NumPy} function \textit{argmax} \cite{harris2020array}. We estimated errors as the uncertainty in identifying the exact positions of the AT peaks on the frequency axis, the exact positions of the AT peaks, considering a 10\% variation of their maximum signal. 
The PSEIT signals (in red) have a dispersive shape, presenting an outer peak and an outer valley. We define $\Delta_{PSEIT-AT}$ as the difference between the frequencies of the outer peak and the outer valley that arise when an MW field is applied.  $\Delta_{PSEIT-AT}$ is obtained by signal processing using the \textit{NumPy} functions \textit{argmax} and \textit{argmin} \cite{harris2020array}. The error in PSEIT measurements is estimated using the same method as for EIT measurements.

\begin{figure}
\includegraphics[width=0.6\columnwidth]{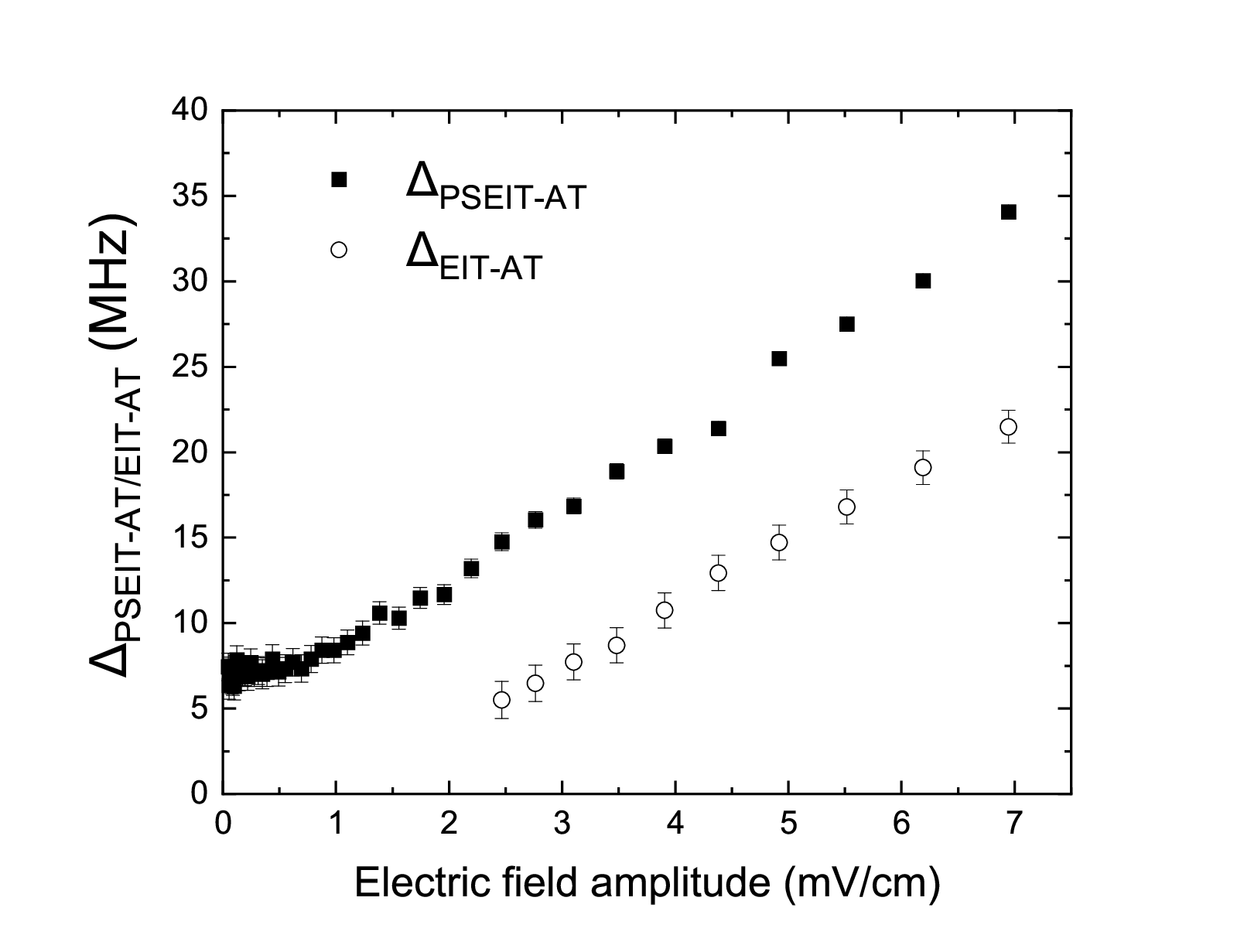}
\caption{$\Delta_{EIT-AT}$ (open circles) and $\Delta_{PSEIT-AT}$ (squares) as functions of the applied electric field. To perform these measurements, the MW power is varied from -7 dBm to -60 dBm.}\label{fig04}
\end{figure}

Figure \ref{fig04} shows the measured $\Delta_{EIT-AT}$ (red circle) and $\Delta_{PSEIT-AT}$ (black square) as a function of the applied MW electric field amplitude. To obtain these measurements, the MW power was varied from -7 dBm to -60 dBm with 1 dBm step; for each power, the spectrum was recorded and the splitting was measured.  It is important to notice that the electric field amplitude is obtained by calibrating the antenna input MW power using $\Delta_{EIT-AT}$ as the MW Rabi frequency is varied, and the $68S_{1/2} \rightarrow 68P_{3/2}$ transition dipole moment \cite{sedlacek2012microwave}. It is also noticeable that the PSEIT technique extends the useful measurement range due to its dispersive shape and larger splitting, reaching a stationary regime at small MW power. The percentage error varied from 5\%(1\%) at high MW field to 20\%(10\%) at low MW field for the EIT(PSEIT) technique.

\begin{figure}
    \centering
    \includegraphics[width=0.6\columnwidth]{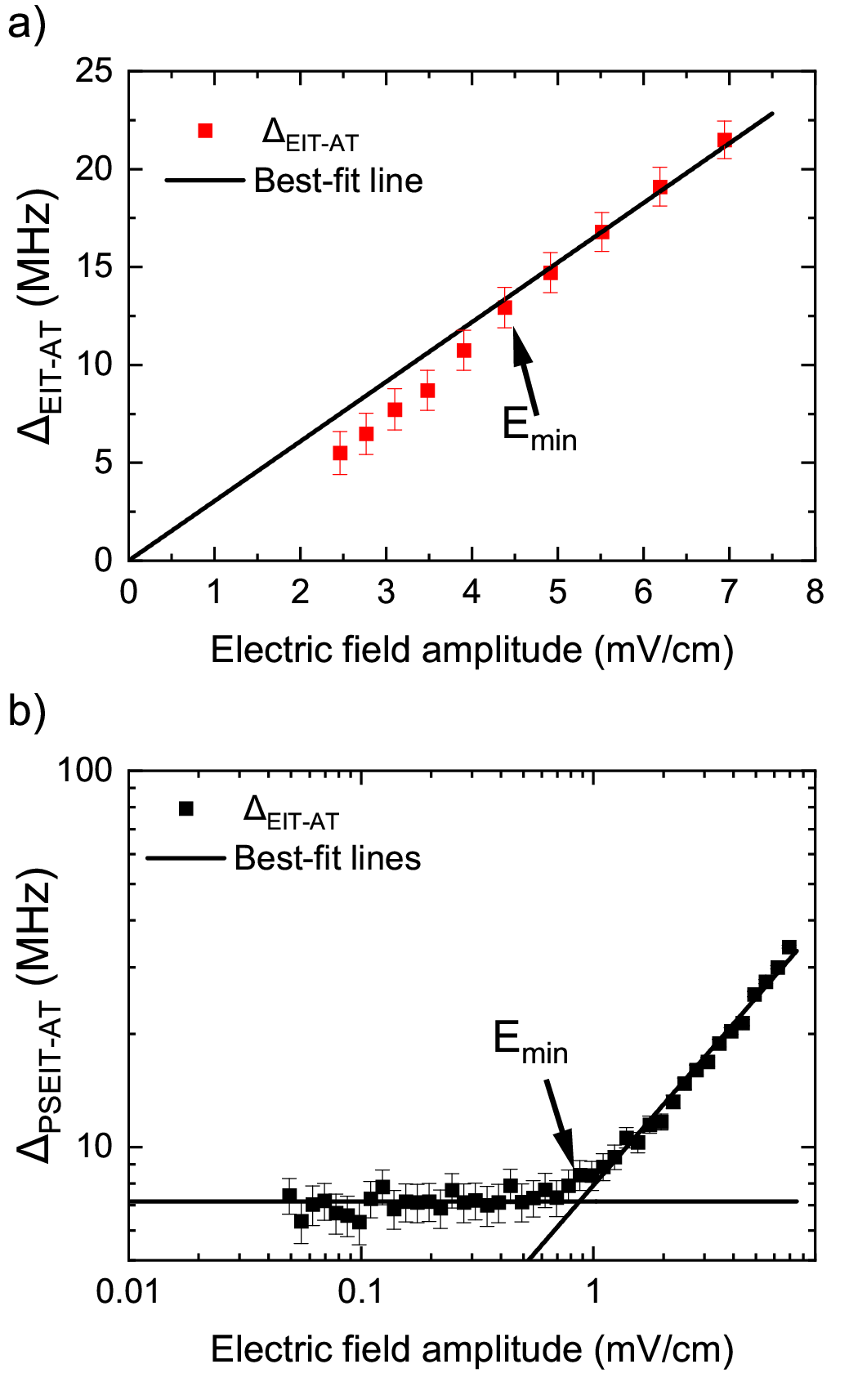}
    \caption{(Color online) Minimum measurable applied MW electric field amplitude for (a) the EIT-AT technique, where the best-fit line is in black and the minimum value is $\approx 4.4$ $\pm$ 0.3 mV/cm, and (b) $\Delta_{PSEIT-AT}$ as a function of the applied electric field on a log-log scale. The intersection between two linear fit functions was taken as the minimum measurable MW electric field amplitude, which is $0.87 \pm 0.10$ mV/cm.}
    \label{fig05}
\end{figure}

It is easy to notice that for smaller fields, $\Delta_{EIT-AT}$ deviates from the linear relation with the MW field, as shown in detail in Fig. \ref{fig05} (a). This nonlinear behavior has been observed \cite{anisimov2011objectively, PhysRevA.90.043849, abi2010interference}, and it has been extensively studied in reference \cite{holloway2017electric}. We have observed that, when fitting the EIT spectrum with two Lorentzian functions with waists $w_{L}$, the measurement is no longer reliable when $w_{L}$ is of the order of $\Delta_{EIT-AT}$. In this situation, the peaks are artificially pulled closer together, reducing the splitting, and the fitting procedure no longer recognizes two independent peaks. A similar effect happens when using the \textit{NumPy} function \textit{argmax}, so we stop the analysis. This is the reason why empty circles stop at higher values of the microwave field. This condition sets a limit for the minimum MW measurable amplitude, which in our case is 4.4 $\pm$ 0.3 mV/cm. It is important to note that such a limit is Rydberg state-dependent. 

For the PSEIT technique, due to the dispersive shape of the signal, the peak and the valley do not pull closer together, allowing a reliable measurement at smaller fields. The minimum MW electric field amplitude was determined from a log-log plot (Fig. \ref{fig05} (b)). For low field,  $\Delta_{PSEIT-AT}$ is constant. As the field increases, $\Delta_{PSEIT-AT}$ increases. The intersection between two linear fit functions was taken as the minimum measurable MW electric field amplitude, which is $0.87 \pm 0.10$ mV/cm, showing $\simeq 5$ times the increase in the sensitivity. Compared to the six-fold increase achieved by Liu \textit{ et al.} \cite{liu2021using}, the enhancement obtained through this technique is comparable and represents an improvement, particularly considering the absence of the need for MW field modulation.

\begin{figure}
    \centering
    \includegraphics[width=0.6\columnwidth]{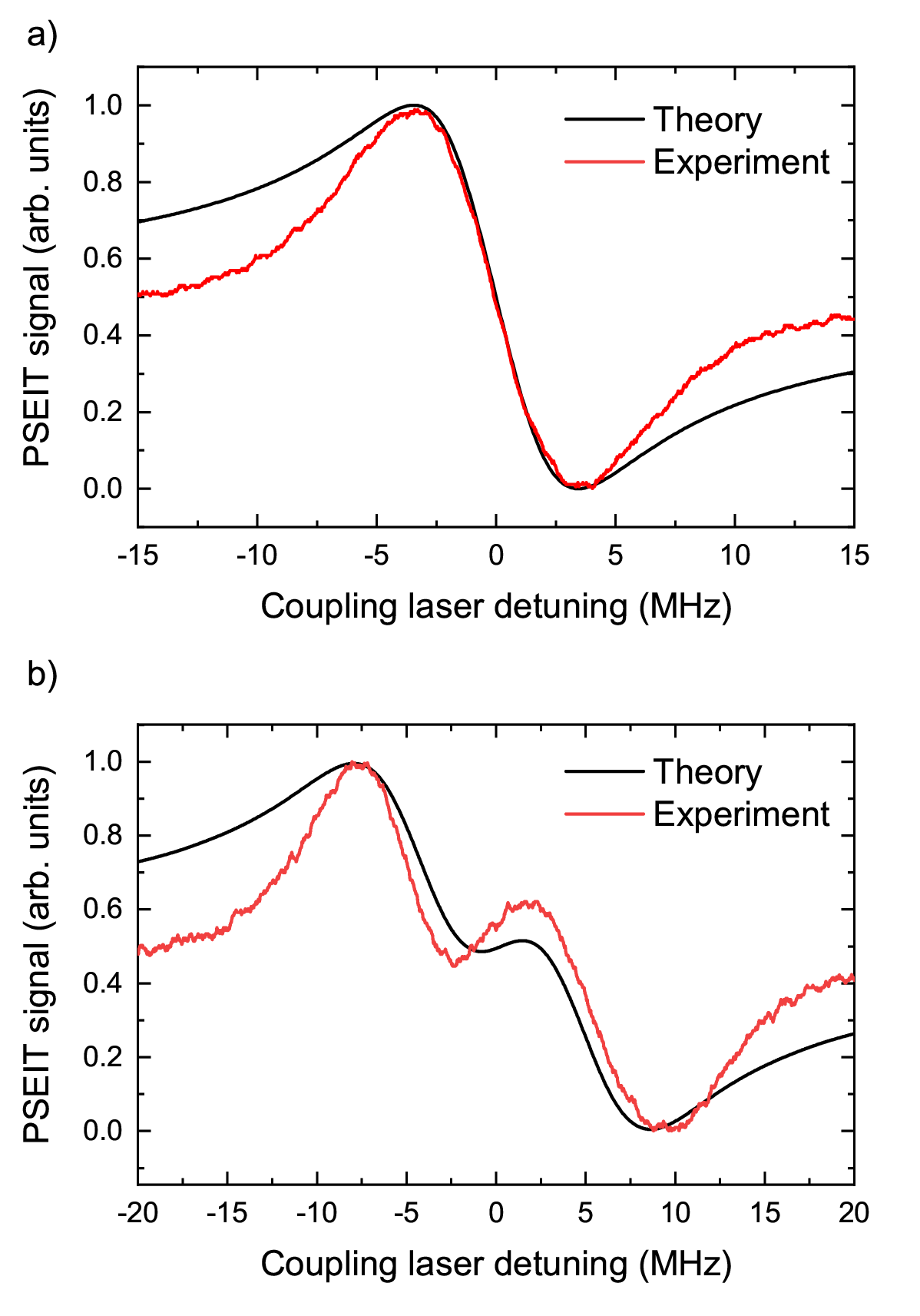}
    \caption{(Color online) Theoretical (black) and experimental (red) transmission spectrum for EIT (a) without applied microwave field, and (b) with a microwave field of 3.10 mV/cm.}
    \label{fig06}
\end{figure}

\begin{figure}
    \centering
    \includegraphics[width=0.6\columnwidth]{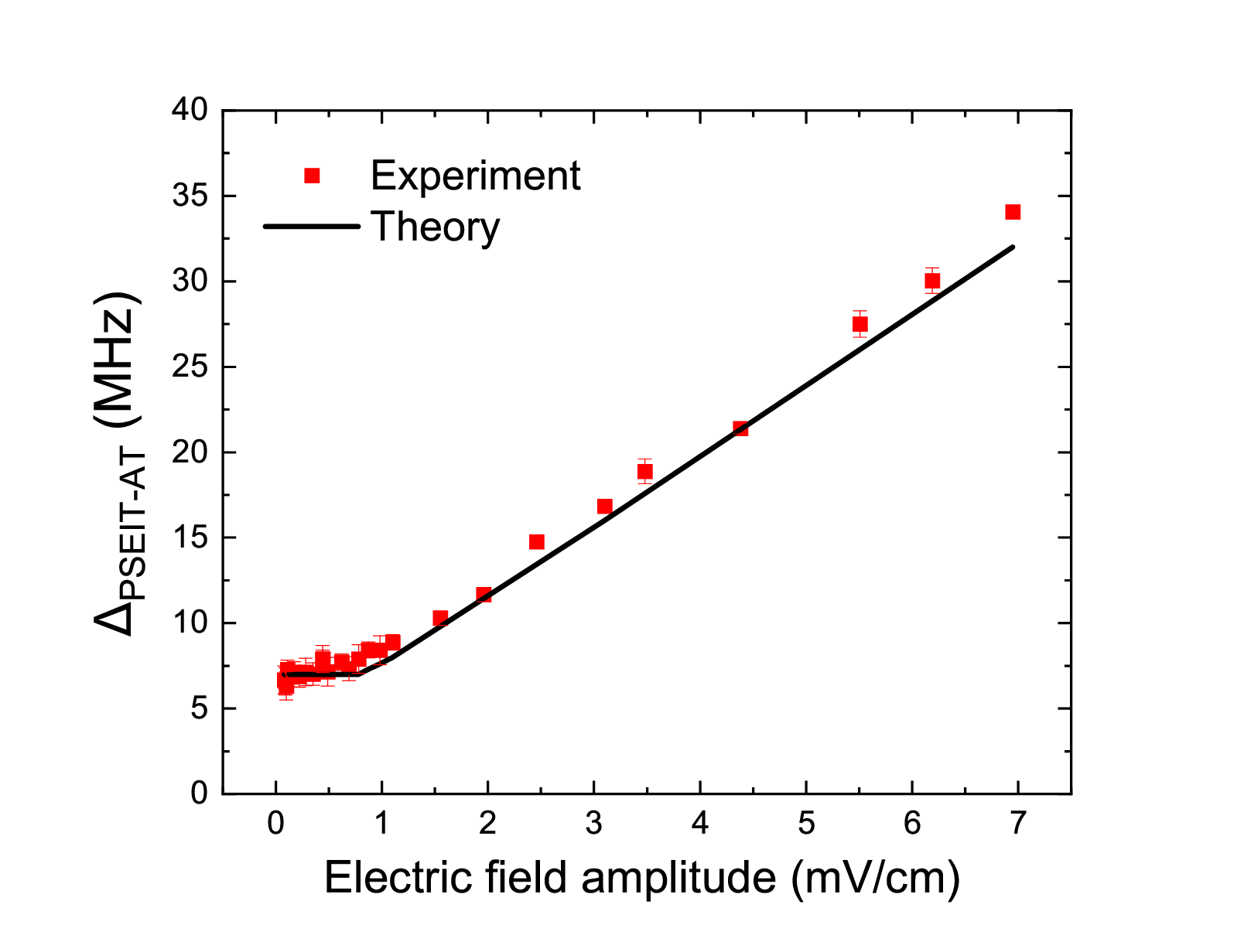}
    \caption{(Color online) $\Delta_{PSEIT-AT}$ as a function of the  MW electric field amplitude. Comparison between experiment (red squares) and theoretical model (line). No fitting parameters.}
    \label{fig07}
\end{figure}

We now compare our experimental and theoretical results. Fig. \ref{fig06} displays theoretical and experimental PSEIT dispersion spectra obtained (a) when no MW field is present, and (b) with an applied MW field intensity of 3.10 mV/cm. Here too, the AT effect is evident, generating a dispersion curve with a larger peak-to-valley distance. It is worth noting that the theoretical PSEIT curves do not precisely match the experimental curves throughout their entire extent. This difference can be attributed to the use of a fictitious dummy state and the lack of clear knowledge regarding the decay rates to that state. Figure \ref{fig07} shows the comparison between the experimental data and the theoretical model for $\Delta_{PSEIT-AT}$ as a function of the amplitude of the MW electric field. Quantitative agreement is remarkably good, considering that there are no adjustable parameters. The model predicts that $\Delta_{PSEIT-AT}$ at zero microwave field depends very weakly on the probe and coupling laser Rabi frequency. Our simulations show that a variation of 50 times in such parameters only produces a change of 30-50\% in $\Delta_{PSEIT-AT}$ at zero microwave field. On the other hand, $\Delta_{PSEIT-AT}$ at zero microwave field is almost linear with the level linewidth $\ket{2}$. If this linewidth can be decreased, the PSEIT limit could be improved.

\subsection{\label{sec:lens-design}Cylindrical lens design and simulation}

In this section, we detail the design and fabrication of the cylindrical MW lens that focuses the microwave field onto the Rydberg atoms. The lens is fabricated in a fused deposition modeling (FDM) three-dimensional (3D) printer with polylactic acid (PLA) plastic, similarly to other 3D printed metalenses and bulk lenses \cite{bulk-printed-lens} with varying infill percentages. The measured PLA relative permittivity is $\varepsilon_r=2.8-i 0.017$ at 11.66 GHz \cite{prb-carac}. We opt for a cylindrical bulk lens design because the large wavelength ($\lambda$) would otherwise require a very thick metalens to achieve complete phase coupling of the wavefront. In addition, as the infill pattern is usually much smaller than $\lambda$, it can be treated as an effective medium whose effective permittivity $\varepsilon_{eff}$ is written as follows \cite{homog-materials}:
\begin{equation}
    \varepsilon_{eff}=1+k_f(\varepsilon_r-1)
\end{equation}
where $k_f$ is the infill percentage. As a result, the effective permittivity is $\varepsilon_{eff}=2.26-i 0.012$, corresponding to an effective refractive index $n=1.5-i 0.004$. The lens curvature radii $R_1$ and $R_2$ define its focal length $f$ as follows \cite{goodman-fourier}:
\begin{equation}
    \frac{1}{f}=(n-1)(\frac{1}{R_1}+\frac{1}{R_2})
\end{equation}
\par We choose $R_1=0.13$ m and $R_2\rightarrow\infty$ (plano-convex lens), resulting in $f=0.26$ m. The lens dimensions are limited by the maximum printing area (17 x 17 cm), resulting in a numerical aperture $NA=0.44$. Thus, the focused beam full width at half maximum (\emph{FWHM}) is limited by Abbe’s diffraction limit \cite{optical-physics-lipson}:
\begin{equation}
    FWHM=\frac{\lambda}{2NA}=2.92\ \text{cm}
\end{equation}
\par
\begin{figure}
\includegraphics[width=0.6\columnwidth]{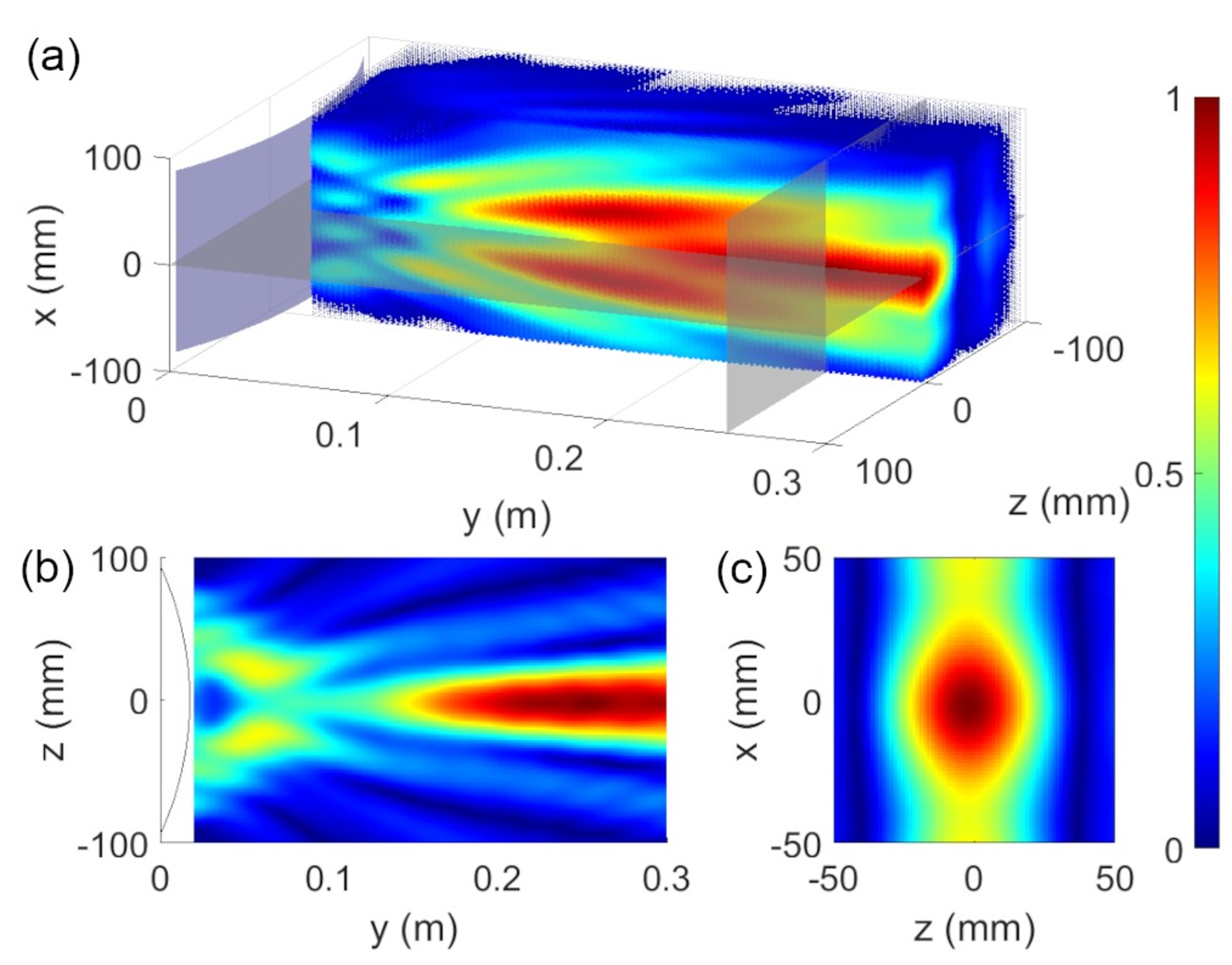}
\caption{\label{fig08}(a) 3D electric field profile of the designed microwave cylindrical lens when excited by a plane wave. At $y=25.5$ cm, the electric field is maximum, and the lens provides a gain of 3.04 times. The lens is represented in the plot by a transparent blue cylinder centered at $x,y,z=0$. Transparent gray planes highlight the $x=0$ longitudinal section (b) and the focus cross-section (c). The beam $FWHM$ is 3.6 cm.}
\end{figure}

 The lens is simulated with a finite element method (COMSOL Multiphysics 6.1) as follows. First, we excite the lens with a plane wave and propagate it for $5~cm$. Next, we use the angular spectrum formalism (ASF) \cite{goodman-fourier} to propagate the diffracted wave for longer distances to reduce the computational cost. The ASF uses Fourier transform algorithms to calculate the diffracted wave producing precise results for the propagation of arbitrary wavefronts through homogeneous media. Figure \ref{fig08}(a) shows the normalized 3D electric field profile of the cylindrical lens (hidden for $z>0$ due to symmetry). Note that there is unwanted diffraction on the top and bottom borders of the lens, due to the large wavelength when compared to the lens aperture. Figure \ref{fig08}(b) shows the longitudinal section of the \textbf{E} field when $x=0$. At $y=25.5$ cm, the electric field \textbf{E} is maximum, and the $FWHM = 3.6$ cm. Its cross-section has an elliptical-like profile as shown in Figure \ref{fig08}(c), with an $FWHM$ higher than the Abbe's limit due to unwanted diffractions that reduce the lens effective numerical aperture. Finally, the gain provided by the lens (i.e., the ratio between the maximum \textbf{E} field over the Rydberg atoms with and without the lens) is 3.04. It should be highlighted that when the lens is excited by a non-plane wavefront, its focal length increases as the wavefront expands; therefore, the exciting horn antenna should be placed as far away from the lens as feasible.

\par
\begin{figure}
\includegraphics[width=0.6\columnwidth]{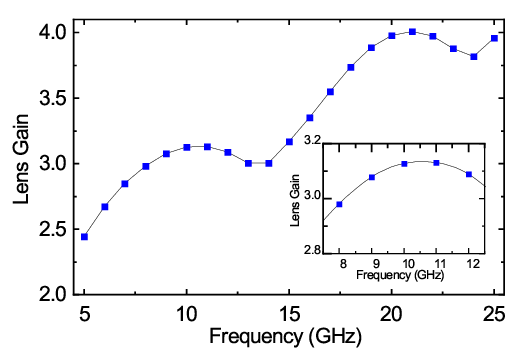}
\caption{\label{fig09} Lens gain as a function of frequency, which is 3 or higher over a very broad frequency range, and its value increases with the operating frequency, owing to the decrease in the width of the focused beam at higher frequencies. Inset zooms in the X-Band, showing an almost flat gain of $3.08 \pm 0.06$.}
\end{figure}

Finally, we assess the operating bandwidth of the microwave lens. Bulk devices typically have broad operating bandwidths, as long as the dielectric is nondispersive in the desired frequency range. To evaluate the operating bandwidth of our CL, we conduct lossless finite-difference time-domain (FDTD) simulations using Ansys Lumerical 2022 R2 to obtain the lens gain from 5 GHz to 25 GHz at $y=25.5$ cm. The results are shown in Fig. \ref{fig09}. It is worth noting that our lens not only spans an exceptionally broad frequency range but also exhibits a gain that increases with frequency, attributable to the narrower beam width in higher frequencies. Furthermore, the gain is almost flat ($3.08 \pm 0.06$) in the X-Band (8 - 12 GHz), as seen in the inset. Note that the gain in the X-Band is slightly higher than the value obtained in COMSOL (3.04) due to the absence of dielectric losses in the frequency-dependent simulation.

\subsection{EIT and PSEIT results with MW lens}\label{lens}

\begin{figure}
    \centering
    \includegraphics[width=0.6\columnwidth]{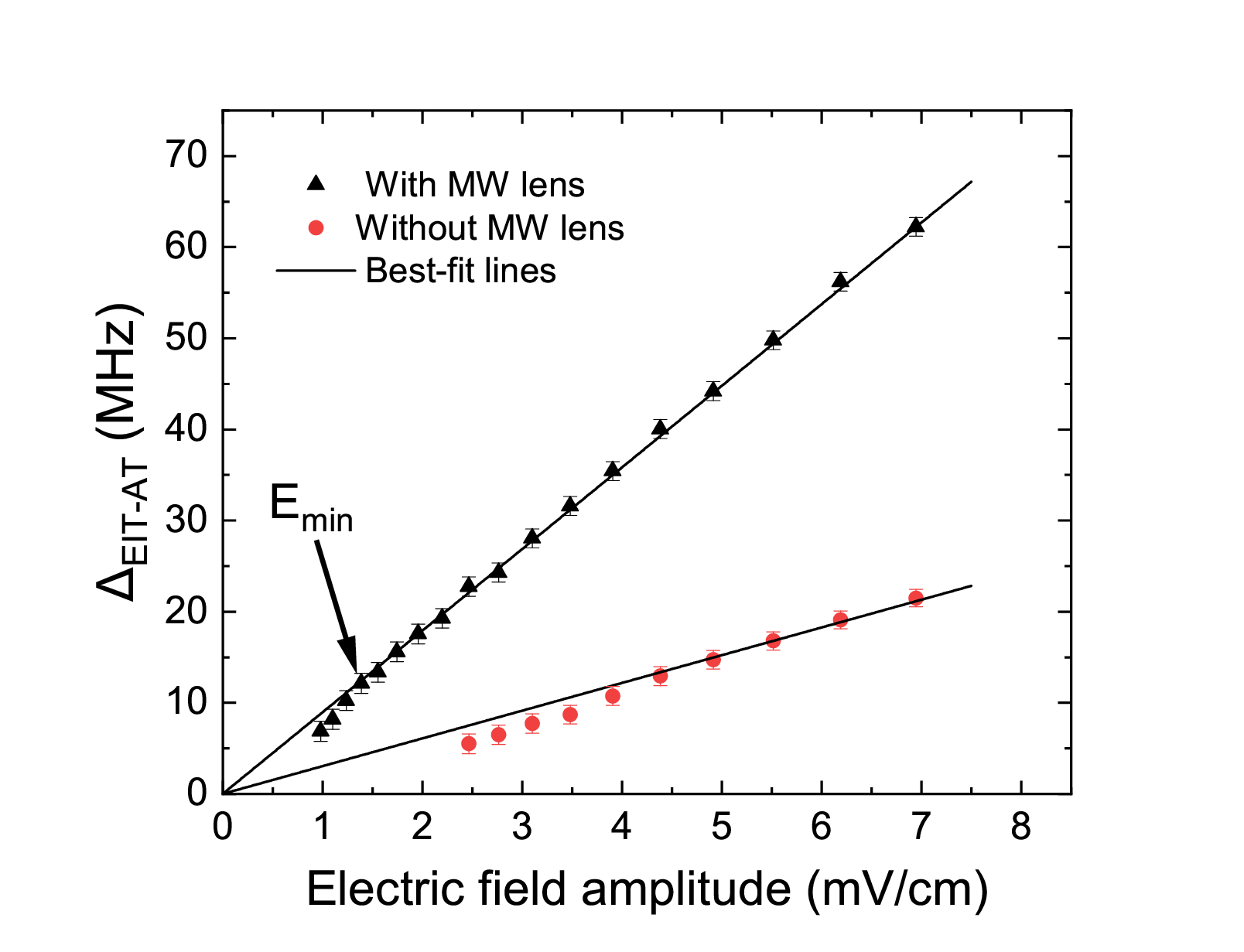}
    \caption{(Color online) $\Delta_{EIT-AT}$ with the lens (black triangles) and the best-fit line (in black), and without the lens (red circles) with the best-fit line (in black). The use of the MW lens extends the measurable region and results in a minimum measurable MW electric field amplitude of $\approx 1.4$ $\pm$ 0.3 mV/cm, which is approximately 3 times smaller than the minimum measured without the lens.}
    \label{fig10}
\end{figure}

In our experimental setup, as depicted in Fig. \ref{fig02}, we placed the MW lens between the microwave horn antenna and the vapor cell. However, due to the non-plane wavefront, the actual focal point of the lens was observed to be slightly displaced from the simulated position, as detailed in Sec. \ref{exp-setup}. The lens was then placed 27 cm away from the vapor cell. According to the lens's simulation, we expect the MW electrical field amplitude to increase by a factor of 3.04 at the vapor cell, which should increase the sensitivity of the techniques. Figure \ref{fig10} shows $\Delta_{EIT-AT}$ as a function of the applied electric field with (black triangles) and without (red circles) the lens. If we apply the same procedure used in Fig. \ref{fig05} (a) for the MW lens, we measure a minimum electric field amplitude of $1.4$ $\pm$ 
 $0.3$ mV/cm. This result indicates that the MW lens has amplified the field in the cell region by a factor of 3, exactly the factor predicted by the model.

\begin{figure}
    \centering
    \includegraphics[width=0.6\columnwidth]{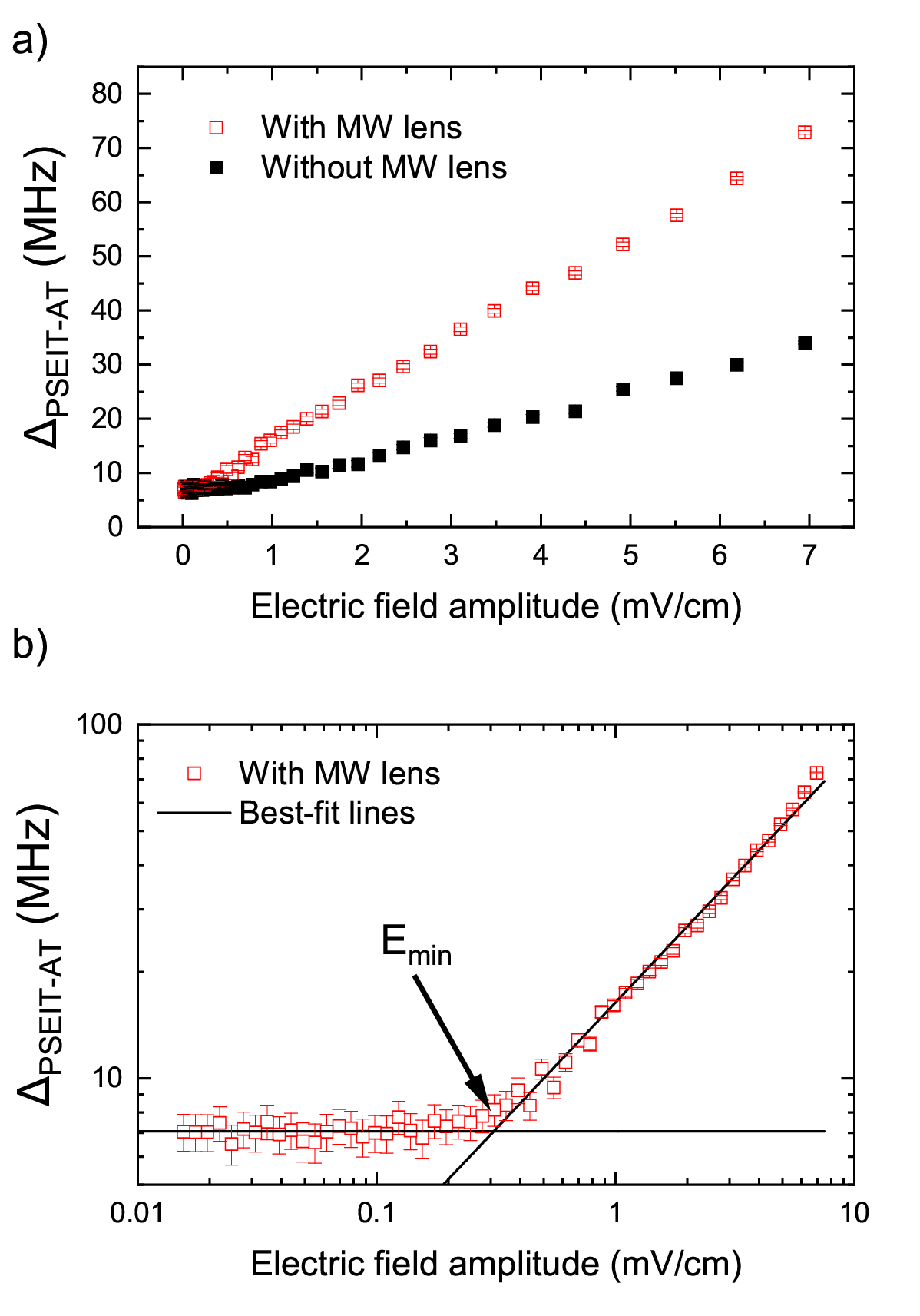}
    \caption{(Color Online) (a) $\Delta_{PSEIT-AT}$ with the lens (open red squares) and without the lens (black squares). The splitting is larger with the lens, but both tend to the same value at low fields. (b) $\Delta_{PSEIT-AT}$ with the lens in a log-log plot. The intersection between two linear fit functions was taken as the minimum measurable MW electric field amplitude, which is $0.31 \pm 0.08$ mV/cm.}
    \label{fig11}
\end{figure}

In the following, we have measured the $\Delta_{PSEIT-AT}$ as a function of the applied electric field with the lens and compared it with the measurements without the lens (Fig. \ref{fig11} (a)). The $\Delta_{PSEIT-AT}$ splitting is larger with the lens, but tends to the same value at low fields as without the lens. Figure \ref{fig11} (b) shows $\Delta_{PSEIT-AT}$ as a function of the applied electric field with the lens in a log-log plot. The intersection between two linear fit functions was taken as the minimum MW electric field amplitude, which is $0.31 \pm 0.08$ mV/cm. Therefore, the PSEIT technique with the lens presents a minimum electric field amplitude $\simeq 15$ times smaller than the EIT technique without the lens.

\section{Conclusion}\label{conclu}
In this work, we have improved the MW field measurement in four-level Rydberg rubidium atoms at a room-temperature vapor cell by employing the PS technique and an MW cylindrical lens. Due to the PSEIT technique, we were able to improve the minimum detectable microwave electric field amplitude by a factor of $\simeq 5$ when compared to regular EIT. The utilization of an MW cylindrical lens resulted in a threefold increase in the MW electric field on the vapor cell, enhancing the minimum detectable microwave electric field by approximately a factor of 3. It is important to note that the lens gain is flat over a broad frequency range. The PSEIT theoretical model reproduces all the main features of the PSEIT spectrum without any adjustable parameters, culminating in quantitative fidelity in predicting both $\Delta_{EIT-AT}$ and $\Delta_{PSEIT-AT}$. Finally, a total improvement factor of $\simeq 15$ was obtained, demonstrating the potential of two simple independent techniques on MW electrometry with Rydberg atoms in a vapor cell. We believe that a 3-photon EIT \cite{JIM3Photon,3photons} could be improved by using the PSEIT technique, since $\Delta_{PSEIT-AT}$ at zero microwave field will also be smaller, allowing the measurement of smaller fields. Moreover, the techniques utilized in this study could be combined with others to enhance sensitivity, such as amplitude modulation of the MW field \cite{liu2021using}.

\section*{Acknowledgement}
This work is supported by grants 2019/10971-0 and 2021/06371-7, S\~{a}o Paulo Research Foundation (FAPESP), and CNPq (305257/2022-6). It was supported by the Army Research Office - Grant Number W911NF-21-1-0211. R.d.J.N. acknowledges support from Funda\c{c}\~ao de Amparo \`a Pesquisa do Estado de S\~ao Paulo (FAPESP), project number 2018/00796-3, and also
from the National Institute of Science and Technology for Quantum Information (CNPq INCT-IQ 465469/2014-0) and the National Council
for Scientific and Technological Development (CNPq).

\bibliography{main}

\end{document}